\documentclass[aps,pre,onecolumn,epsf,graphicx]{revtex4}
\usepackage{epsfig,latexsym,amssymb}

\begin{document}

\title{\bf\noindent Path integrals for stiff polymers  
applied to membrane physics}

\author{D.S. Dean$^{(1)}$ and R.R. Horgan$^{(2)}$}

\affiliation{
(1) IRSAMC, Laboratoire de Physique Th\'eorique, Universit\'e Paul Sabatier, 
118 route de Narbonne, 31062 Toulouse Cedex 04, France\\
(2) DAMTP, CMS, University of Cambridge, Cambridge, CB3 0WA, UK\\
e-mail:dean@irsamc.ups-tlse.fr, rrh@damtp.cam.ac.uk}
\date{11 April 2007}
\begin{abstract}
Path integrals similar to those describing stiff polymers arise in the
Helfrich model for membranes. We show how these types of path
integrals can be evaluated and apply our results to study the
thermodynamics of a minority stripe phase in a bulk membrane. The
fluctuation induced contribution to the line tension between the
stripe and the bulk phase is computed, as well as the effective
interaction between the two phases in the tensionless case where the
two phases have differing bending rigidities.
\end{abstract}  
\maketitle
\vspace{.2cm}
\pagenumbering{arabic}
\section{Introduction}
Recently their has been much interest in the effective interactions
between components of membranes of different composition. These
effective interactions can be direct basic interactions such as
electrostatic and van der Waals forces. However the fact that the
membrane fluctuates also leads to effective interactions which are due
to how the different components, or inclusions, alter the membrane
fluctuations.  Coarse grained models, based on the Helfrich
model\cite{hel,boal}, of multicomponent membrane describes the
membrane in terms local mechanical properties, for instance the
bending rigidity $\kappa_b$, the Gaussian rigidity $\kappa_g$ or the
spontaneous curvature \cite{leibler,tani,saxena,misbah2,netz,nepi,dema}.  
The effective, fluctuation mediated, interaction between regions of differing
rigidity can be computed using a cumulant expansion giving the
effective pair-wise component of the interaction between two
regions. This term is of order $\delta\kappa_{b/g}^2$ and is the
analogue of the pair-wise component of van der Waals forces. However
when $|\delta\kappa_{b/g}|$ is large then this two-body, or dilute,
approximation will break down and a full $N$-body calculation is
needed. We should expect the dilute approximation to break down
reasonably frequently as experimentally measured values of $\kappa_b$
for commonly occurring lipid types vary from $3$ to $30k_BT$.

\begin{figure}
\epsfxsize=0.5\hsize \epsfbox{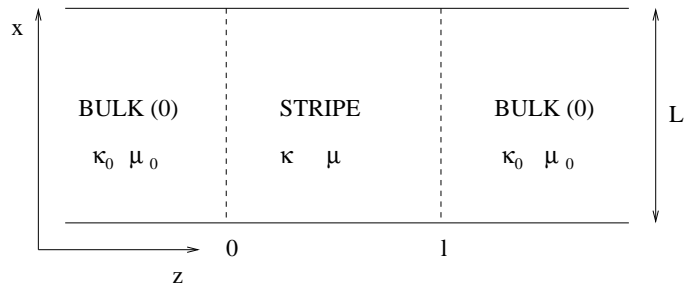}
\caption{Schematic diagram of striped membrane configuration with mechanical parameters
(rigidity and elasticity indicated. }
\label{fpol}
\end{figure}

In this paper we show how the full $N$-body calculation for a system
with spatially varying rigidity and elasticity, can be carried out for
stripe geometries of the type shown in Figure \ref{fpol}. Within this geometry
we can compute the contribution to the line tension between the two
phases due to membrane fluctuations. This can be seen as a
renormalization of the line tension already present due to basic
interactions such as van der Waals, electrostatic and steric forces.
In addition, in some cases, we can evaluate the effective interaction
between the two interfaces as a function of their separation $l$.

They key point in our calculation is that we convert the usual
functional integral into a path integral where the direction in which
the physical parameters change is treated like a fictitious time
variable within the path integral formalism. The authors have already
applied this approach to electrostatic problems where it has proved to
be efficient for carrying out computations for films \cite{film},
interfaces \cite{st} and in cylindrical geometries such as
lipid-tubules \cite{tube}.

The paper is organized as follows. In section \ref{model} we describe
the model and show how it can be analyzed using path integrals which are
mathematically identical to those arising for stiff polymers.
In section \ref{bulk-stripe} we present the results of our computations. We start 
with an analysis of bulk homogeneous membranes and show how some standard results can be
recovered using our path integral formalism and then how the method can be applied to
a striped membrane system. The formalism is then used to compute the membrane 
fluctuation--induced contribution to the line tension between two phases. In section 
\ref{applications} the fluctuation--induced Casimir force in a striped system is calculated for typical
physical situations. In section \ref{conclusion} we conclude with a discussion of our results. 
The generalized Pauli-van Vleck formula used to evaluate the path integrals 
is described in detail in Appendix \ref{PvV}. Some aspects of this
approach have been previously described in the literature
\cite{pik,pi1,pi2}, however our approach is slightly different and
self-contained. Hence for the sake of completeness (and because the
results seem to be relatively unknown) we include this detailed
description of the approach.

\section{\label{model} The model}

In the Monge gauge the Helfrich model for a membrane whose height fluctuations
are denoted by $h$ is given by
\begin{equation}
H = {1\over 2} \int_{A_p} d^2{\bf x}\ \left[ \kappa \left(\nabla^2
h\right)^2 + \mu \left(\nabla h\right)^2\right].\label{hel}
\end{equation}
In this model we have neglected the Gaussian rigidity $\kappa_g$ and
so for ease of notation we denote the bending rigidity simply by
$\kappa$. In this form of the Helfrich model there is no spontaneous
curvature and it is implicitly assumed that the fluctuations of $h$
are small. The term $\mu$ can be interpreted as a local elastic or
surface energy. When $\mu$ is constant it can be interpreted as a
surface tension. The integral in ${\bf x} = (z,y)$ in Equation (\ref{hel})
is over the projected are of the membrane $A_p$. The physical area of
the membrane is larger than $A_p$ due to fluctuations and we denote it
by $A = A_p +\Delta A$, where $\Delta A$ is the excess
area due to fluctuations. In the limit of small height fluctuations
(i.e. to quadratic order in $h$) the excess area is given by
\begin{equation}
  \Delta A = {1\over 2}\int_{A_p} d^2{\bf x} \ \left(\nabla h\right)^2.
\end{equation}
The canonical partition function can be written as a functional integral
over the height field $h$
\begin{equation}
Z = \int d[h] \exp(-\beta H),
\end{equation}
where $\beta = 1/k_BT$, $T$ is the canonical temperature and 
$k_B$ Boltzmann's constant.

If the mechanical parameters $\kappa$ and $\mu$ only vary with the
coordinate $z$ then we can express $h$ in terms of its Fourier
decomposition in the direction $y$ writing
\begin{equation}
h(z,y) = {1\over \sqrt{L}}\sum_k {\tilde h}(z,k) \exp(iky).
\end{equation}
We have imposed periodic boundary conditions in the $y$ direction and thus have
$k = 2\pi n/L$, where $L$ is the width of the system and $n$ is an integer. 
Note that we are assuming that the interfaces between differing phases are
straight and and thus lie  at constant values of $z$.
This is a realistic assumption 
if the bare (in the absence of height fluctuations) line tension $\gamma_0$
is positive and large. We will in fact see that the renormalization
of the line tension $\gamma$ between phases due to the height fluctuations is
positive and thus this assumption remains valid (and is in fact reinforced)
by height fluctuations.   

In the limit of large $L$ the sum over modes can be written as
\begin{equation}
\sum_k \to L\ \int {dk\over 2\pi}\ .
\end{equation}
We now find that the Hamiltonian decomposes as
\begin{equation}
H = \sum_k H_k,
\end{equation}   
where
\begin{equation}
H_k = {1\over 2} \int dz \kappa(z) \left( 
{\partial^2 {\tilde h}(z,-k)\over  \partial z^2}{\partial^2 {\tilde h}(z,k)\over  \partial z^2}\right) + (2\kappa(z) k^2 + \mu(z)) \left( 
{\partial {\tilde h}(z,-k)\over  \partial z}{\partial {\tilde h}(z,k)\over  \partial z}\right) + (\mu(z) k^2 + \kappa(z)) {\tilde h}(z,k){\tilde h}(z,-k).
\end{equation}
The field $h$ is real and so we have the relation ${\overline {\tilde h}(z,k)}= {\tilde h}(z,-k)$. 
The full functional integral for the partition function $Z$ can then be written as 
\begin{equation}
Z = \prod_{k\geq 0}\Theta_k,
\end{equation} 
where
\begin{equation}
\Theta_k = \int d[X] \exp\left(-{1\over 2} \int_0^l dt\left[
a_2(k,t)  \left({d^2 X\over dt^2}\right)^2 + a_1(k,t) \left({d X\over dt}\right)^2
+ a_0(k,t)^2 X_t^2   \right]\right).
\end{equation}
In the above, the coefficients are given by
\begin{eqnarray}
a_2(k,t) &=& \beta\kappa(t) \nonumber \\
a_1(k,t) &=& \beta(2\kappa(t) k^2 + \mu(t)) \nonumber \\
a_0(k,t) &=& \beta(\mu(t) k^2 + \kappa(t)),
\end{eqnarray}
and $l$ is the length of the system (in the $z$ direction). The above
path integral is that arising in an elastic model of a semi-flexible
polymer \cite{semi} (in one dimension) where  $\kappa =a_2/\beta$ is the
rigidity of the polymer, the term proportional to $a_1$ represents the
elastic energy and the term proportional to $a_0$  represents an
external harmonic potential.  In this model the length of the polymer
is not fixed, in contrast to the worm-like chain model where the
magnitude of the tangent vector is fixed. The method of evaluation of the
above type of path integral is given in Appendix \ref{PvV} and we find that 
it takes the form
\begin{eqnarray}
 K({\bf X}, {\bf Y};t) &=&  (2\pi)^{-{N\over 2}}\left[ \det\left(B(t))\right)\right]^{{1\over 2}} \nonumber \\
&&\exp\left(-{1\over 2}{\bf X}^T A_I(t) {\bf X} -{1\over 2} {\bf Y}^T A_F(t) {\bf Y}
+ {\bf X}^T B(t) {\bf Y}\right),
\end{eqnarray}
where the initial condition vector is ${\bf X} = (X, U)$ with $X=
X(0)$ and $U = dX/ds|_{s=0}$, and the final condition vector is ${\bf
Y} = (Y, V)$ with $Y= X(t)$ and $V = dX/ds|_{s=t}$. When the
coefficients $a_k$ are independent of $t$ the classical action can be
written as a combination of surface terms (using the equation of
motion) as
\begin{equation}
S_{cl}({\bf X}, {\bf Y}) = {1\over 2} \left( a_2\left[{dX\over ds}{d^2X\over ds^2}\right]_0^t
- a_2 \left[X {d^3 X\over ds^3}\right]_0^t  + a_1 \left[X{dX\over ds}\right]_0^t\right).
\end{equation}
The above expression is in general rather complicated but can be used
to determine the matrices $A_F$, $A_I$ and $B$. Also, when the
coefficients are independent of time, the time reversed trajectories
have the same weight as the original trajectories. Thus the path
integral going from $(X,U)$ to $(Y,V)$ in time $t$ has the same value
of the path integral going from $(Y,-V)$ to $(X,-U)$ in time
$t$. Mathematically this means that
\begin{equation}
A_F = S A_I S, \label{aiaf}
\end{equation}
and 
\begin{equation}
B^T = S B S,
\end{equation}
where
\begin{equation}
S = \pmatrix{{1\ \ \ 0}\cr {0 \ -1}}.
\end{equation}
In the limit of large $t$ the propagator $K$ should factorize, 
as it is dominated by the lowest eigenfunction, and so we 
should find that $B(t) \to 0$ as $t\to\infty$. This can be verified by 
explicit calculation in the cases considered here.
\section{\label{bulk-stripe} Calculations for bulk and striped systems}

To start with we will show how the path integral formalism introduced
here reproduces some standard results concerning bulk systems. We
consider a bulk system of projected length $l$ and projected width
$L$; we thus have a projected area $A_p = Ll$.  Periodic boundary
conditions are imposed in both directions $z$ and $y$.  The free
energy is given by
\begin{equation}
F = -k_B T \sum_k \ln(\Theta_k),
\end{equation}
where
\begin{equation} 
\Theta_k = \int d{\bf X}\, K_k({\bf X}, {\bf X}, l).
\end{equation}
Using Equation (\ref{eqpvv}) and the notation developed in the appendix we
find
\begin{equation} 
\Theta_k = \det\left(B(k,l))\right)^{1\over 2} \det\left(A_I(k,l) +
A_F(k,l) -2 B(k,l)\right)^{-{1\over 2}}.
\end{equation}
The classical equation of motion in this case has solutions
\begin{equation}
X(t) = a \cosh(p t) + b \sinh(pt) + c \cosh(q t) + d \sinh(qt), \label{Xpq}
\end{equation}
where
\begin{equation}
p = k, \label{eqp}
\end{equation}
and
\begin{equation}
q = (k^2 + m^2)^{1\over 2},\label{eqq}
\end{equation}
where we have defined
\begin{equation}
m^2 = {\mu\over \kappa},
\end{equation}  
and so $m$ is an inverse length scale.  The expressions for $A_F$ $A_I$
and $B$ can be computed using computer algebra but they simplify in
the (thermodynamic) limit $l\to \infty$.  We define
\begin{equation}
A^*_{I/F} = \lim_{l\to\infty} A_{I/F},
\end{equation}
and find that
\begin{equation}
A^*_I = \beta \kappa \pmatrix{{pq (p+q)\ \ \  pq} \cr {\ \ \ pq \ \ \ \ \ \  \ \ (p+q)}}. \label{eqAI}
\end{equation}
We also find that for large $l$  $B(l) \to 0$, and 
\begin{equation}
\det\left(B(l)\right) \approx pq (p+q)^2 \exp\left(-(p+q) l\right).
\end{equation}

The extensive part of the bulk free energy is thus
\begin{equation}
{F\over A_p} = {k_B T\over 2\pi}\int_{\pi\over L}^{\pi\over a}
dk \left[ k + (k^2 + m^2)^{1\over 2}\right],
\end{equation}
where we have introduced the ultra-violet cut-off length scale $a$
which corresponds to the lipid size.  The infra-red cut-off scale
(where needed) is given by $L$ (the lateral size of the system). The
excess area of the system is given by
\begin{equation}
\Delta A = {\partial F\over \partial \mu},
\end{equation}
and in the tensionless limit where $\mu = 0$ we find the well known result
\begin{equation}
{\Delta A \over A_p} = {k_BT\over 4\pi \kappa}\ln\left({L \over a}\right).
\end{equation}  
In the case where $\mu\neq 0$ we find that
\begin{equation}
{\Delta A \over A_p} = {k_BT\over 4\pi \kappa}\sinh^{-1}\left({\pi \over a m}\right).
\end{equation}
This gives
\begin{equation}
{\Delta A \over A_p}
\approx {k_BT\over 4\pi \kappa}\ln\left({2\pi \over a m}\right),
\end{equation} 
when $a\ll 1/m$.

For a striped geometry where the length of the bulk phase is $l_0$, and
large and that of the minority phase is $l$ we find that for this
composite striped (hence the superscript $s$ in what follow) system we
have, as $l_0\to \infty$,
\begin{eqnarray}
\Theta^{(s)}_k(l,l_0) &=& \int  d{\bf X} d{\bf Y} K_k({\bf X}, {\bf Y},l)
K_k^{(0)}({\bf X}, {\bf Y},l_0) \nonumber \\
&=& \left[\det(B^{(0)}(l_0))\right]^{1\over 2}\left[\det(B(l))\right]^{1\over 2}
\left[\det\left(A_F^{(0)*} +A_I(l)\right) \right]^{-{1\over 2}}\nonumber \\
&& \left[\det\left(A_I^{(0)*} +A_F(l)
-B^T(l)( A_F^{(0)*} + A_I(l))^{-1}B(l)\right)\right]^{-{1\over 2}}, \label{theta}
\end{eqnarray}
where the superscript $(0)$ refers to the bulk phase and the absence
of this superscript refers to the minority phase. In the limit $l\to
\infty$ the above expression simplifies giving
\begin{equation}
\Theta^{(s)}_k(l,l_0) \approx \left[\det(B^{(0)}(l_0))\right]^{1\over 2}
\left[\det(B(l))\right]^{1\over 2} \left[\det(A_I^{(0)*} +A_F^*)\right]^{-1},
\end{equation}
where we have used Equation (\ref{aiaf}). In order to compute the free energy
cost of the interface between the two phases we subtract the separate bulk free energies
for large systems of size $l_0$, corresponding to the bulk, and of size $l$, corresponding
to the minority phase, from that of a large striped system composed of length $l_0$ of the
bulk phase and $l$ of the minority phase. This free--energy difference is
\begin{equation}
\Delta F = -k_B T\sum_k \ln\left({\Theta^{(s)}_k(l,l_0)\over 
 \Theta^{(0)}_k(l_0)  \Theta_k(l)}\right),
\end{equation}  
which gives
\begin{equation}
\Delta F = -{k_B T\over 2} \sum_k \ln\left({\det(A_I^{(0)*} +A_F^{(0)*})
\det(A_I^* +A_F^*)\over \det(A_I^{(0)*} +A_F^*)^2}\right).\label{sum1}
\end{equation}
The above expression is in general quite complicated but when $\kappa$
and $\kappa_0$ are non-zero, then at large $k$ the eigenvalues $q$ 
(given by Equation (\ref{eqq})) in the stripe phase becomes asymptotically
equal to $q_0$, the corresponding eigenvalue in the bulk phase.  We
thus find that the sum in Equation (\ref{sum1}) is ultra-violet divergent
and is dominated by the term
\begin{equation}
\Delta F = {L k_B T\over a}\ln\left({1-\Delta^2/4 \over 1 -\Delta^2}\right),
\label{lenergy}
\end{equation}
where
\begin{equation}
\Delta = {\kappa-\kappa_0\over \kappa+\kappa_0}.
\end{equation}
We may interpret this result as the existence of a height fluctuation induced
line tension
$\gamma_{hf}$ between the two phases (note the factor of a half as there
are two interfaces) given by
\begin{equation}
\gamma_{hf} = { k_B T\over 2 a}\ln\left({1-\Delta^2/4 \over 1 -\Delta^2} \right).
\end{equation}
Thus the dominant contribution to the fluctuation induced line tension
between the two phases comes from the miss--match in their bending
rigidities.  We also remark that it does not depend on $m$ and is only
dependent on $\kappa$ and $\kappa_0$ through $\Delta^2$, which is a
symmetric function of the two rigidities. As an example the
fluctuation induced line tension between two phases whose rigidity
differs by a factor of $10$ has an energy of about $0.5\ k_B T$ per
lipid at the interface.

\begin{figure}
\epsfxsize=0.5\hsize \epsfbox{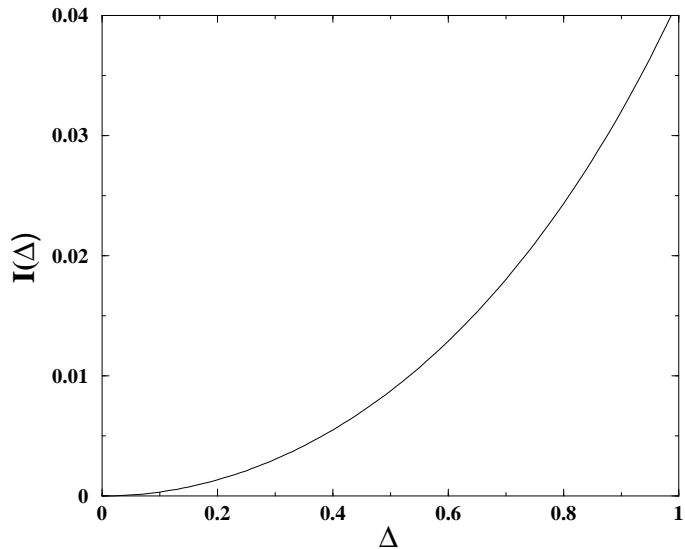}
\caption{The function $I(\Delta)$ defined in Equation (\ref{gcorrection}), 
where $\Delta = {(\kappa-\kappa_0)/ (\kappa+\kappa_0)}$.
Note that $I(\Delta)=I(-\Delta)$.} \label{I}
\end{figure}

The correction terms to $\gamma_{hf}$ are UV convergent and, after some
manipulation, $\gamma$ can be expressed as a power series in $ma/\pi$
as
\begin{equation}
\gamma_{hf} = \frac{k_B T}{a}\left[\frac{1}{2}\ln\left({1-\Delta^2/4 \over 1 -\Delta^2} \right)
+\frac{ma}{\pi}\;I(\Delta) - \frac{1}{128}\;\frac{\Delta^2}{(1-\Delta^2/4)}\left(\frac{1}{3}\left(\frac{ma}{\pi}\right)^4
-\frac{1}{5}\left(\frac{ma}{\pi}\right)^6\right)\right]\,. \label{gcorrection}
\end{equation}
$I(\Delta)$ is shown in Figure \ref{I} for $\Delta>0$ and is a
non-negative function of $\Delta$ with $I(0)=0$ and
$I(\Delta)=I(-\Delta)$. From Figure \ref{I} we see that $I(\Delta)$
has a maximum value $I(\infty) \sim 0.04$. For $\mu = 10^{-2}N/m$,
$\kappa = 25k_BT \sim 10^{-19}J$ and $a = 10^{-9}m$, we find
$ma=\sqrt(\mu/\kappa)a \sim 0.32$ and so $ma/\pi \sim 0.1$. The
correction to the leading term due to non-zero $\mu$ is thus expected
to be certainly less than $O(1\%)$.

\section{\label{applications} Applications}
In this section we discuss the application of the theory developed
above to two cases in the system with a stripe as shown in Figure
\ref{fpol}. These cases are distinguished by the values of the masses
in the two regions, $m_0=\sqrt(\mu_0/\kappa_0)$,
$m=\sqrt(\mu/\kappa)$, which control the relationship of the surface
to bending energies. In general, the boundary conditions satisfied by
the system will vary and will determine the precise way in which our
formalism is applied and the form taken by the relevant Helfrich
action. We consider two cases that might be thought of as extreme
situations and are chosen to show how the results are markedly
different depending on the exact situation.  We concentrate on
computing the Casimir force across the stripe which can be interpreted
as a force between the opposing interfaces. The Casimir free energy,
$F_C(l)$, is therefore normalized to $F=0$ in the limit $l \to
\infty$. We have
\begin{equation}
F_C(l)~=~F(l,l_0)~-~\lim_{l \to \infty}F(l,l_0)\left.\right|_{l+l_0=\mbox{constant}}\,,
\end{equation}
where
\begin{equation}
F(l,l_0)~=~-k_BT\sum_k \ln \left(\Theta_k^{(s)}(l,l_0)\right)\;,
\end{equation}
and $\Theta_k^{(s)}(l,l_0)$ is defined in Equation (\ref{theta}). It is understood that the total volume of
the system is held fixed by imposing $l+l_0=\mbox{constant}$, and that $l_0$ is large compared with
any system-specific length scale.
                                                                                                                                               
\subsection{${\bf m_0=m=0}$}
This case corresponds to an untethered membrane which is tensionless, as is the case for a membrane in the presence of the various lipid species in solution.
Upon a change in the physical area of the membrane, $A$, lipid molecules can
leave or enter meaning that any area change costs no free energy. 

Then we have $\mu_0=\mu=0$. In this case, the choice for the general solution to the classical
equations of motion is not given by Equation (\ref{Xpq}) but by
\begin{equation}
X(t) = a \cosh(p t) + b \sinh(pt) + c\:t \cosh(p t) + d\:t \sinh(pt)\;. \label{Xpp}
\end{equation}
The method follows the manipulations of section \ref{bulk-stripe}, and appendix \ref{PvV}. We
find that
\begin{equation}
F_C(l,\Delta)~=~\frac{k_BT}{2}\sum_k\ln \left(1+a_2(l,\Delta)e^{-2kl}+a_4(l,\Delta)e^{-4kl}\right)\;,
\end{equation}
where
\begin{eqnarray}
a_2(l,\Delta)&=&\frac{\Delta^2}{(1-\Delta^2/4)}\left[k^2l^2
\left(\frac{1-\Delta/2}{1+\Delta/2}\right)-\frac{3}{2}\right]\;, \nonumber \\
a_4(l,\Delta)&=&\frac{9\Delta^4}{16(1-\Delta^2/4)^2}\;.
\end{eqnarray}

\begin{figure}
\epsfxsize=0.5\hsize \epsfbox{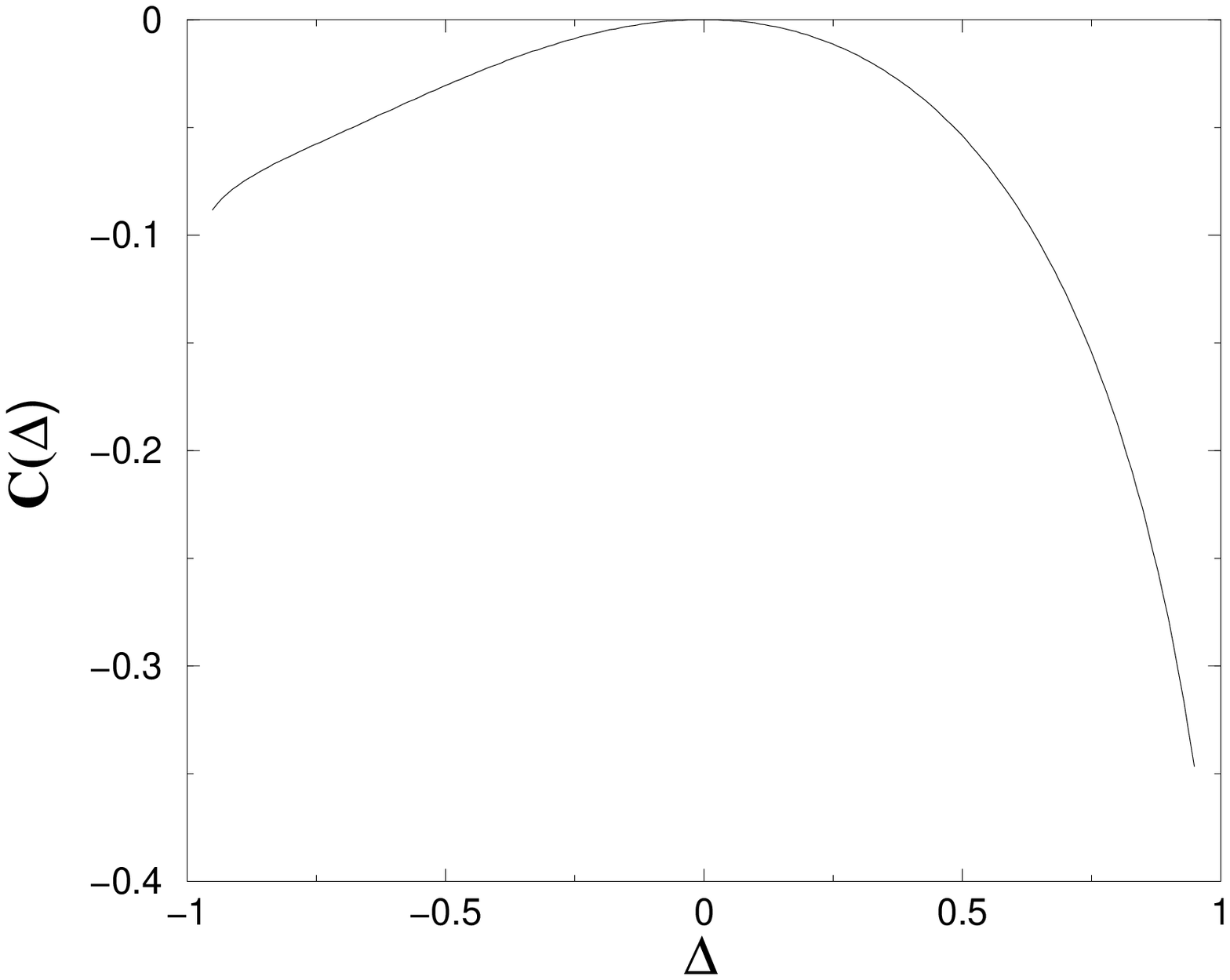}
\caption{The function $C(\Delta)$ defined in Equation (\ref{FCzero}), 
where $\Delta = {(\kappa-\kappa_0)/ (\kappa+\kappa_0)}$.
Note that $C(\Delta)\ne C(-\Delta)$.} \label{C}
\end{figure}

We note that all terms are invariant under $\Delta \to -\Delta$ except
the first term in $a_2(l)$, and hence the Casimir free energy 
is not invariant under this transformation in this case. On
dimensional grounds we have
\begin{equation}
F_C(l,\Delta)~=~\frac{C(\Delta)k_BT}{l}\;, \label{FCzero}
\end{equation}
and $C(\Delta)$ is shown in Figure \ref{C}. Since $C(\Delta)<0$ for $\Delta \ne 0$, the Casimir force 
is attractive and is given by
\begin{equation}
f_C(l,\Delta)~=~-\frac{\partial}{\partial l}\frac{C(\Delta)k_BT}{l}~=~\frac{C(\Delta)k_BT}{l^2}\;,
\end{equation}
with $|C(\Delta)| \lessapprox 0.4$.

\begin{figure}
\epsfxsize=0.5\hsize \epsfbox{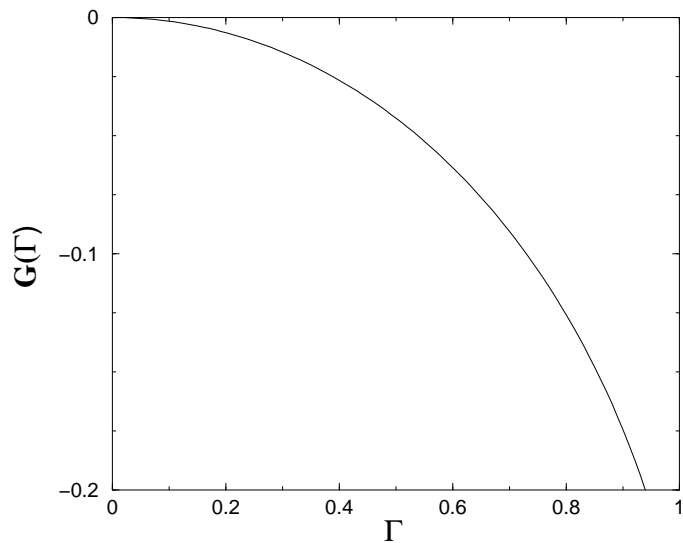}
\caption{The function $G(\Gamma)$ defined in Equation (\ref{FGnonzero}),
where $\Gamma = {(\mu-\mu_0)/ (\mu+\mu_0)}$.
Note that $G(\Gamma) = G(-\Gamma)$.} \label{G}
\end{figure}

\subsection{$\bf m_0,m > 0$}
This case is for non-zero $\mu_0$ and $\mu$. The projected area 
is constant but any change in the area results in a free energy change. 
In practice, the typical observed values of $\mu$ give $ma \sim 0.3$ (see section
\ref{bulk-stripe}) and consequently the width of the stripe, $l$,
satisfies $l \ll 1/m$. As shown in section \ref{bulk-stripe} the free
energy has a contribution that corresponds to the energy of the
interfaces between the stripe and the bulk medium and includes the UV
divergent part of the sum over modes. The remaining, $l$-dependent,
terms are UV convergent and are cut-off but exponential factors on a
scale $k \lessapprox 1/l$. Thus, the Casimir free energy excluding the
interface energy gets contributions only from low mode number $k \ll
1/m$ and so we can approximate the eigenvalues in Equations (\ref{eqp}) and
(\ref{eqq}) by
\begin{equation}
p~=~k\;,~~~~q~=~m\;.
\end{equation}
Also, only leading terms in $e^{ql}$ will survive; the others will be
suppressed by factors of $e^{-ml}$. Following the derivation of the
previous subsection, we find a result that is independent of
$\kappa_0$ and $\kappa$, as one might expect on dimensional grounds for $\mu
\gg \kappa/l^2$. We get
\begin{equation}
F_G(l,\Gamma)~=~\frac{k_BT}{2}\sum_k\ln \left(1-\Gamma^2e^{-2kl}\right)\;,
\end{equation}
where $\Gamma = (\mu-\mu_0)/(\mu+\mu_0)$. The leading corrections are
suppressed by the factor $(1/ml)^2$. For the values of $m$ considered
this is of order $(a/l)^2 \sim 1/N^2$, where $N$ is the number of
lipid molecules across the strip. Similarly to before we may write
\begin{equation}
F_G(l,\Gamma)~=~\frac{G(\Gamma)k_BT}{l} + O\left(\frac{1}{(ml)^2}\right)\;. \label{FGnonzero}
\end{equation}
$G(\Gamma)$ is symmetric under $\Gamma \to -\Gamma$, and is shown in
Figure \ref{G}. We see that $|G(\Gamma)| \lessapprox 0.2$ and, being
negative for $\Gamma \ne 0$, gives rise to an attractive Casimir force
of maximum magnitude $0.2k_BT/l^2$.

The more general case where $m \sim 1/l$ is very much more complicated
and the expressions cannot be presented here but need to be
investigated computationally in the three parameter $\mu,\kappa,l$
space. We have verified that this analysis is feasible but complicated
and so we have chosen not to present it in this paper. However, any
more general result for the Casimir force will interpolate between the
two extremes presented here and so we would expect a leading
contribution to behave like $-ck_BT/l^2$ with $c \lessapprox 1$.

\section{\label{conclusion}Conclusion}

In this paper we have shown how the formalism developed in earlier
work \cite{tube} can be applied to the more general case of higher
derivative Gaussian energy functions such as apply to the path
integral analysis of stiff polymers and the Helfrich model for
membranes. Some aspects of this approach have been previously
described in the literature \cite{pik,pi1,pi2}. However, our approach
is slightly different and self-contained. Hence, for the sake of
completeness (and because the results seem to be relatively unknown)
we have included a detailed description of this approach. In
particular, we have shown how to generalize the Pauli-van Vleck
formula for the evolution kernel of all theories of this type.

As a model system we have considered a toroidal lipid membrane with
one very large circumference and the other finite of length $L$, with
a stripe of width $l$ wrapped around the finite circumference and of
different, minority, lipid type to the bulk, majority, type. This is
shown schematically in Figure \ref{fpol}. This geometry imposes
periodic boundary conditions on the system.  We have shown how to
compute, in general, two important energies in this system, namely the
energy, or line tension, associated with the lipid-lipid interface 
and the Casimir force between the interfaces, as a function of width $l$. An important
controlling parameter, $m$, has the dimensions of a mass and is given
by $m^2=\mu/\kappa$. We have presented these calculations explicitly
for the cases where $m=0$ and $ma \sim 1,~l \gg 1/m$, where $a$ is the
inter-lipid spacing. These correspond, respectively, to the cases
where the actual area, $A$, or the projected area, $A_p$, is
conserved. In the latter case, we calculate the mean excess area of the
system $\Delta A/A_p$ in section \ref{bulk-stripe}.  In both cases,
the interface energy is positive and the Casimir force attractive as
can be seen from Equations (\ref{lenergy}),(\ref{FCzero}), (\ref{FGnonzero})
and the associated figures.  Our general result is that in appropriate
dimensionful units the energy coefficient is $ck_BT$ where, at maximum,
$c \sim 1$. This is to be compared with the natural bending rigidity
with lies in the range $5k_BT \le \kappa \le 100k_BT$.

The general case where $l \sim 1/m$ is complicated and long, and
although we have the results we have not presented an explicit
analysis in the $\mu,\kappa,l$ parameter space because of the
complexity.  However, there is no computational or algebraic
impediment to carrying this out.

In terms of relevance to the physical system we might consider two
scenarios in the two-lipid model discussed here. Either the minority
lipid can be dissolved in the majority lipid to form a homogeneous
phase for the mixture, or the minority lipid can precipitate out of
solution and form a pure minority phase, the stripe in our idealized
case, within the pure majority phase. Which situation is stable is, of
course, decided by a competition between the free energies of the
configurations which is in turn dependent on the boundary conditions
imposed. However, it is clear that the attractive Casimir force will
tend to reduce the stripe width $l$, presumably by evaporation of
minority lipid from the interface into solution. The interface energy
is constant throughout such a process but will always tend to minimize
the interface length. A stability analysis, however, requires a
computation of the free energy of the mixed phase which our
calculation does not address. However, as has been discussed in 
\cite{dema}, the suppression of lipid mode fluctuations by
confining the membrane in a stack will change the free-energy of both
configurations and so can affect their stability; an effect which can
be analyzed by our methods.

\appendix
\section{\label{PvV} The generalized Pauli-van Vleck formula}

In this appendix we show how generalized quadratic path integrals can
be evaluated giving a generalization of the Pauli-van Vleck formula.
The treatment is very close to that of \cite{pi2} and is based
on the Chapman-Kolmogorov decomposition of the path integral. 
We consider the following path integral
\begin{equation}
K({\bf X},{\bf Y};t) = \int_{{\bf X}(0)={\bf X}}^{{\bf X}(t) ={\bf  Y}}d[X]\ \exp\left( -S[X]\right),
\end{equation}
where $S$ is a quadratic action which will have the general form
\begin{equation}
S[X] = {1\over 2} \int_0^tds\;\sum_{k=0}^N a_k \left(d^k X\over ds^k\right)^2.
\end{equation} 
In general the coefficients $a_k$ can be time dependent but for the
problems related to membranes studied here we will only require the
results for $a_k$ constant. The usual Wiener measure occurring in path
integrals has $N=1$ and the corresponding path integral is that for
standard Brownian motion, or a free particle, with $a_0=0$. If $a_0 \neq
0$ in this case, then the path integral corresponds to that of a simple harmonic
oscillator with $a_0 = m\omega^2$ and $a_1 = m$ thus relating the
coefficients $a_i,~i=0,1$ to the mass $m$ and frequency of the
oscillator. The path integrals arising in section \ref{model} are, of
course, for the case $N=2$ which, as mentioned above, also arises for
the path integrals of stiff or semi-flexible polymers. We also refer
the reader to the approach of \cite{pik} which is based on an
eigenfunction expansion method for the case $N=2$.

Now one must state how the initial and final points of the path
integral should be specified. The presence of the term $\left(d^N
X/ds^N\right)^2$ means that the paths that contribute to the path
integral are ones where the derivatives $d^{N-1} X/ds^{N-1}$ and lower
must be continuous. The path integral should therefore be specified in
terms of the vector ${\bf X} = (X, X^{(1)}, X^{(2)} \cdots X^{(N-1)})$
where we have used the notation $X^{(k)} = d^k X/ds^k$. We can now
decompose the path integral using the Chapman-Kolmogorov formula
\begin{equation}
K({\bf X},{\bf Z};t+t') = 
\int d{\bf Y} K({\bf X},{\bf Y};t)K({\bf Y},{\bf Z};t').\label{ck}
\end{equation}
This decomposition ensures the continuity of the path $X(t)$ up to
its $N-1$th derivative. 

The classical path is given by the one that minimizes the action:  
\begin{equation} 
{\delta S\over \delta X(s)} = 0,
\end{equation}
with the boundary conditions on the end points ${\bf X}(0) = {\bf X}$
and ${\bf X}(t) = {\bf Y}$. This gives a total of $2N$ boundary
conditions ($N$ from each end). The equation for the classical path
can be written as
\begin{equation}
\int ds'  {\delta^2 S\over \delta X(s)\delta X(s')}X_{cl}(s') = 0,
\end{equation}
which is a linear differential equation of order $2N$. For instance, when the 
$a_k$ are constant it reads  
\begin{equation}
\sum_{k=0}^{N} (-1)^k a_k {d^{2k}\over dt^{2k}} X_{cl}(s) = 0.
\end{equation}
There are thus $2N$ linearly independent solutions to this equation
and their coefficients are linearly related to the $2N$ conditions for
the end points.  The classical action is a quadratic form in the
initial and final condition vectors ${\bf X}$ and ${\bf Y}$ and we can
write
\begin{equation}
S_{cl}({\bf X}, {\bf Y}) = {1\over 2}\left[{\bf X}^T A_I(t) {\bf X} + {\bf Y}^T A_F(t) {\bf Y}
-2 {\bf X}^T B(t) {\bf Y}\right],
\end{equation}
where we have used the subscripts $I$ and $F$ to denote the initial
and final coordinates. We now write the path $X(s)$ as
$X(s) = X_{cl}(s) + x(s)$, where the boundary conditions imply that
${\bf x}(0) ={\bf 0}$ and ${\bf x}(t) = {\bf 0}$. The path integral
can now be written as
\begin{eqnarray}
K({\bf X}, {\bf Y};t) &=& \exp\left(- S_{cl}({\bf X}, {\bf Y})\right)
\int_{\bf 0}^{\bf 0} d[x]\exp\left(-{1\over 2} \int_0^t ds ds'
x(s'){\delta^2 S\over \delta X(s)\delta X(s')}x(s)\right) \nonumber \\
&=&  Q(t)\exp\left(- S_{cl}({\bf X}, {\bf Y})\right),\label{eqfor}
\end{eqnarray}
where we formally write can write 
\begin{equation}
Q(t) = 
\det\left( {\delta^2 S\over \delta X(s)\delta X(s')}\right)^{-{1\over 2}},~~~~0 \le s,s' \le t.   
\end{equation}
The above functional determinant can be evaluated using an
eigenfunction expansion, however for higher order operators this
quickly becomes impractical.  Instead, we return to the Chapman
Kolmogorov formula Equation (\ref{ck}) and pursue its consequences using
the formal result Equation (\ref{eqfor}). Explicitly carrying out
the intermediate integration over ${\bf Z}$, we find that
\begin{eqnarray}
K({\bf X}, {\bf Z};t+t') &=& (2\pi)^{{N\over 2}} Q(t) Q(t')
\det\left(A_I(t) + A_F(t')\right)^{-{1\over 2}} \times \nonumber \\
&&\exp\left(-{1\over 2} {\bf X}^T \left[A_I(t) - B(t) (A_I(t') + A_F(t))^{-1}
B^T(t) \right]{\bf X}\right) \times \nonumber \\
&&\exp\left(-{1\over 2} {\bf Z}^T \left[A_F(t') - B^T(t') (A_I(t') + A_F(t))^{-1}
B(t') \right]{\bf Z}\right) \nonumber \times \\
&& \exp\left( {\bf X}^T B(t)   (A_I(t') + A_F(t))^{-1}B(t') {\bf Z}\right),
\end{eqnarray}
Now comparing the quadratic forms and prefactors we find the following
relations:
\begin{eqnarray}
A_I(t+t') &=& A_I(t) - B(t)(A_I(t') + A_F(t))^{-1} B^T(t) \\
A_F(t+t') &=& A_F(t') - B^T(t')(A_I(t') + A_F(t))^{-1} B(t')\\
B(t+t') &=& B(t) (A_I(t') + A_F(t))^{-1} B(t') \label{eqB}\\
Q(t+t') &=& (2\pi)^{N\over 2} Q(t) Q(t') \det\left(A_I(t) + A_F(t')\right)^{-{1\over 2}}.\label{eqQ}
\end{eqnarray} 
In \cite{pi2} it is pointed out that relation Equation (\ref{eqQ}) above can
be used to derive a differential equation for $Q$. However, a more rapid way
of finding $Q$ is to note that taking the determinant of both sides of 
Equation (\ref{eqB}) gives
\begin{equation}
\det\left(B(t+t')\right) = \det\left(B(t)\right) \det\left(B(t')\right)
 \det\left(A_I(t) + A_F(t')\right)^{-1},
\end{equation}
and using this relation we find that by direct substitution into
Equation (\ref{eqQ}) that the solution for $Q$ is
\begin{equation}
Q(t) = (2\pi)^{-{N\over 2}}\left[ \det\left(B(t)\right)\right]^{{1\over 2}}.
\end{equation}
The generalized form of the Pauli-van Vleck formula may thus be written in
the familiar form (for $N=1$ and its generalization to higher dimensions) 
\begin{equation} 
K({\bf X}, {\bf Y};t) = (2\pi)^{-{N\over 2}}\det\left[
\partial S_{cl}\over \partial  X_i \partial Y_j\right]^{1\over 2}
\exp\left(-S_{cl}({\bf X}, {\bf Y})\right).
\label{eqpvv}
\end{equation}

\pagestyle{plain}

\end{document}